\documentclass[11pt]{article}
\usepackage[letterpaper,left=1in,right=1in,top=0.85in,bottom=0.85in,headheight=15pt]{geometry}
\usepackage[T1]{fontenc}
\usepackage[utf8]{inputenc}
\usepackage{amsmath,amssymb}
\usepackage{newtxtext,newtxmath}
\usepackage{microtype,graphicx,caption,setspace,booktabs,tabularx,longtable}
\usepackage[mathlines]{lineno}
\usepackage{xcolor,fancyhdr}
\usepackage[backend=biber,style=nature,sorting=none,maxbibnames=5,minbibnames=1,doi=true,isbn=false,url=false]{biblatex}
\usepackage[hidelinks]{hyperref}
\usepackage{xurl}
\let\cite\supercite
\newcommand{\EIO}{\ensuremath{\mathrm{Eu_2Ir_2O_7}}}
\newcommand{\TTO}{\ensuremath{\mathrm{Tb_2Ti_2O_7}}}
\newcommand{\DTO}{\ensuremath{\mathrm{Dy_2Ti_2O_7}}}
\graphicspath{{figures/}}

\newcommand{\papertitle}{
Transducing quantum-spin-ice correlations into Weyl Fermi-arc transport at a synthetic Kondo lattice interface
}
\title{\bfseries\fontsize{18}{21}\selectfont\papertitle}
\author{%
Tsung-Chi Wu$^{1,2,3,*}$, Michael Terilli$^{1}$, Christian Zaprianov$^{1}$, Eun Sang Choi$^{4}$,\\ David Graf$^{4}$, Qinghua Zhang$^{5}$,  Lin Gu$^{6}$ and Mikhail Kareev$^{1}$\\[0.8em]
\small
$^{1}$Department of Physics and Astronomy, Rutgers University, Piscataway, New Jersey 08854, USA\\
\small
$^{2}$Department of Materials Science and Engineering, University of California, Berkeley, Berkeley, CA 94720, USA\\
\small
$^{3}$Department of Materials Science and NanoEngineering, Rice University, Houston, Texas 77005, USA\\
\small
\small
$^{4}$National High Magnetic Field Laboratory, Tallahassee, FL 32310, USA\\
\small
$^{5}$Beijing National Laboratory for Condensed Matter Physics, Institute of Physics, \\
\small
Chinese Academy of Sciences, Beijing 100190, China\\
\small
$^{6}$Beijing National Center for Electron Microscopy and Laboratory of Advanced Materials,\\ 
\small
Department of Materials Science and Engineering, Tsinghua University, Beijing 100084, China\\
\small
\small
\small
$^{*}$To whom correspondence should be addressed:\\
\small
\href{mailto:tcwu@physics.rutgers.edu}{tcwu@physics.rutgers.edu} (T.-C.W.); 
}

\date{}
\begin{document}
\doublespacing
\maketitle
\begingroup\bfseries
Quantum spin ice hosts collective excitations and quantum fluctuations that are largely electrically silent because the materials supporting them are insulating. Coupling this magnetic environment to surface-localized Weyl states offers a route to encode its correlations directly into Fermi-arc transport. Experimental access to this regime, however, has remained largely unexplored. Here, we realize a synthetic Kondo lattice interface between the Weyl semimetal $\mathrm{Eu_2Ir_2O_7}$ and the quantum-spin-ice candidate $\mathrm{Tb_2Ti_2O_7}$, and demonstrate that quantum-spin-ice correlations are imprinted on Weyl Fermi-arc transport. At ultra-low temperature, electronic transport evolves from a sixfold anisotropic response at low field into a twelvefold response over an intermediate field window, before entering a reentrant sixfold anisotropy at high field. No analogous twelvefold response occurs at the classical-spin-ice interface. The quantum--classical contrast shows that the same itinerant electronic channel acquires qualitatively different symmetry information when interacting with the quantum-fluctuating Tb moments, consistent with an interfacial coupling between the Weyl surface states and the transverse, multipolar degrees of freedom, in addition to the longitudinal, dipolar contribution. Our work establishes synthetic Kondo lattice interfaces as a route for transducing frustrated quantum magnetism with multipole correlations into distinct Weyl Fermi-arc transport responses.
\par\endgroup
\newpage
\subsection*{Introduction}
The interface between topological semimetals and quantum spin liquids provides a setting in which itinerant quasiparticles can couple directly to emergent gauge fields and fractionalized magnetic excitations.\cite{Hermele2004,Gingras2014,Balents2010,Savary2017} Specifically, Weyl semimetals host massless chiral quasiparticles and surface Fermi arcs,\cite{Wan2011,Armitage2018,Burkov2016,Liu2021} whereas quantum spin ice provides a three-dimensional setting in which frustrated moments can support emergent gauge correlations and fractionalized excitations.\cite{Hermele2004,Gingras2014,Balents2010,Savary2017} Bringing these two dissimilar materials into direct contact therefore creates the possibility to examine how itinerant Weyl fermions respond to a fluctuating quantum-magnetic environment and whether that response can reveal degrees of freedom absent in the classical limit.\cite{Armitage2018,Gingras2014,Rau2019} Theory already suggests that coupling itinerant electrons to quantum-spin-ice fluctuations can qualitatively reconstruct electronic interactions, including unconventional pairing and broken electronic symmetries.\cite{She2017,RauKee2014,Wang2016}

Bulk probes such as neutron scattering, optical spectroscopy, thermodynamics, and local magnetometry are powerful for identifying magnetic correlations and excitations, but they do not directly report how those correlations are encoded in interfacial electronic transport.\cite{Gardner1999,Gardner2010,Gingras2014,Rau2019,Petit2012,Guitteny2015} A Weyl surface channel provides a complementary electronic probe: since Fermi-arc states are localized at the surface and possess a strongly momentum-dependent electronic structure, changes in the neighboring magnetic environment can be converted into detectable changes in transport anisotropy.\cite{Wan2011,Resta2018,Armitage2018,Burkov2016,Liu2021} This principle was established in epitaxial $\EIO/\DTO$ heterostructures, where the field evolution of classical, predominantly dipolar spin-ice correlations was imprinted onto the $\EIO$ electronic response; related iridate/titanate heterostructures have likewise demonstrated electrical sensitivity to ice-rule breaking.\cite{Wu2025,Kareev2025,Zhang2023,Xing2025} Classical spin ice itself is well established as a dipolar, ice-rule-constrained magnetic state with emergent defects and characteristic field responses.\cite{denHertog2000,Bramwell2001,Castelnovo2008,Gardner2010,Rau2015} The experimental challenge is therefore no longer whether interfacial transport can access dipolar spin-ice magnetism, but whether it can resolve the additional non-dipolar, quantum-fluctuating degrees of freedom.\cite{Onoda2011,Gingras2014,Rau2019,Savary2017}

$\TTO$ provides such a test system. The Tb$^{3+}$ ground-state manifold is a non-Kramers doublet in which the longitudinal pseudospin component carries an Ising magnetic dipole, whereas the transverse components encode electric-multipolar degrees of freedom.\cite{Onoda2011,Ruminy2016,Rau2019,Gingras2014,Gingras2000,Kao2003} In $\TTO$, the strongly field-dependent magnetic response involves low-lying crystal-field excitations, magnetoelastic coupling, and quadrupolar correlations, while also depending on stoichiometry and thickness.\cite{Molavian2007,Rule2006,Cao2008,Ruff2010,Sazonov2010,Sazonov2013,Fennell2014,Ruminy2019,Turrini2021,Taniguchi2013,Takatsu2016,Gritsenko2020,Kadowaki2022,Roll2024,Petit2012,Guitteny2015,Yasui2002,Bovo2017TTOFilm,Wen2021PyrochloreFilms,Anderson2024TTOFilms,Han2025LSMOTTO} Neutron-scattering and thermodynamic studies have established multiple field-induced reorganizations of the Tb correlations whose character depends sensitively on field magnitude and crystallographic direction, together with evolution toward high-field polarized configurations.\cite{Rule2006,Cao2008,Ruff2010,Sazonov2010,Sazonov2013,Petit2012,Guitteny2015,Roll2024} Replacing $\DTO$ with $\TTO$ in the heterostructures therefore does more than change the rare-earth ion. Instead, it changes the interfacial local-moment environment from a predominantly classical dipolar spin-ice sector to a quantum non-Kramers environment that includes both longitudinal dipolar and transverse multipole degrees of freedom.\cite{denHertog2000,Bramwell2001,Castelnovo2008,Onoda2011,Gingras2014,Rau2019} The resulting $\EIO/\TTO$ interface therefore realizes a synthetic two-dimensional frustrated Kondo lattice geometry in which Weyl fermions are exchange-coupled to quantum-fluctuating Tb moments.\cite{Machida2007,Machida2010,RauKee2014,Kavai2021}

Here, we report on the experimental realization of atomically engineered $\EIO/\TTO$ heterostructures and demonstrate that this synthetic Kondo lattice interface encodes the quantum-spin-ice correlations into Weyl Fermi-arc transport. At ultralow temperatures, the in-plane angular magnetoresistance evolves from a sixfold response at low field to a twelvefold response over an intermediate field range, then reenters the sixfold anisotropy at high field. No analogous twelvefold reconstruction occurs in the $\EIO/\DTO$ classical-spin-ice reference. The quantum--classical contrast shows that the same itinerant electronic channel acquires qualitatively different symmetry information when coupled to the quantum-fluctuating Tb moments, consistent with interfacial coupling between the Weyl surface states and the transverse, multipolar degrees of freedom, in addition to the longitudinal, dipolar contribution.\cite{Onoda2011,Rau2019,She2017,RauKee2014} The twelvefold anisotropy therefore provides an electronic fingerprint of multipolar correlations in quantum spin ice that has no counterpart in classical spin ice.\cite{denHertog2000,Bramwell2001,Gardner2010,Rau2015}

\subsection*{A synthetic Kondo lattice interface between the Weyl semimetal and quantum spin ice}
\begin{figure}[p]
\centering
\includegraphics[width=\textwidth]{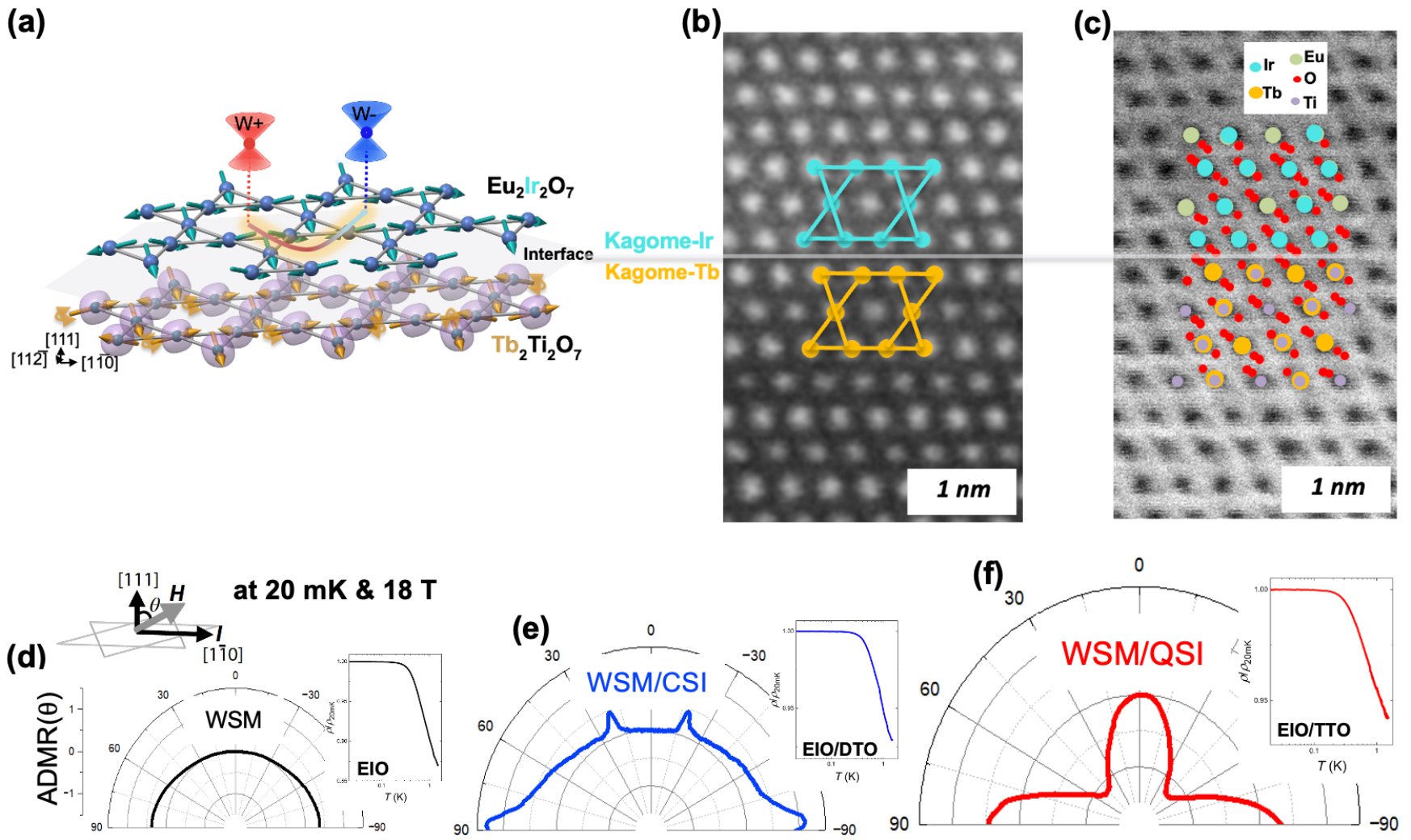}
\caption{\textbf{A synthetic Kondo lattice interface between the Weyl semimetal and quantum spin ice.} \textbf{a}, Schematic of the (111)-oriented $\EIO/\TTO$ heterointerface. Surface-localized Weyl/Fermi-arc states of $\EIO$ form the itinerant electronic subsystem\cite{Wan2011,Liu2021,Armitage2018}, while the adjacent non-Kramers Tb kagome layer provides the frustrated local-moment subsystem; the interface therefore realizes an atomically defined two-dimensional Kondo-coupled geometry. \textbf{b}, High-angle annular dark-field scanning transmission electron microscopy (HAADF-STEM) atomic-column map of the pyrochlore iridate/titanate interface along the $[1\bar{1}0]$ projection, with the Ir and Tb magnetic-sublattice motifs highlighted; scale bar, 1~nm. \textbf{c}, Annular bright-field scanning transmission electron microscopy (ABF-STEM) image resolving the oxygen sublattice along the same projection; the schematic overlay indicates the $\EIO/\TTO$ interfacial stacking sequence for clarity; scale bar, 1~nm. \textbf{d--f}, Out-of-plane angular magnetoresistance at 20~mK and 18~T for bare $\EIO$ (\textbf{d}), the classical-spin-ice interface $\EIO/\DTO$ (\textbf{e}), and the quantum-spin-ice interface $\EIO/\TTO$ (\textbf{f}) in the indicated rotation geometry. Insets show normalized temperature-dependent resistivity. $\textrm{ADMR}(\theta)=[\rho_{xx}(\theta,H)-\rho_{xx}(0^{\circ},H)]/\rho_{xx}(0^{\circ},H)$, with $H$ rotating away from [111]  to [1$\bar{\text{1}}$0].
}
\label{fig:fig1}
\end{figure}

Figure~\ref{fig:fig1}a shows the materials architecture of a synthetic Kondo lattice interface between the Weyl semimetal $\EIO$ and quantum spin ice candidate $\TTO$.\cite{Wan2011,Liu2021,Kareev2025}. The (111)-oriented $\EIO/\TTO$ junction places kagome-derived Weyl Fermi-arc surface states in direct proximity to a kagome layer of non-Kramers Tb local moments (Fig.~\ref{fig:fig1}a).\cite{Kareev2025,Onoda2011,Ruminy2016} In this sense, the interface realizes a two-dimensional frustrated Kondo lattice geometry, where itinerant Weyl fermions are exchange-coupled across an atomically defined boundary to a lattice of frustrated rare-earth moments.\cite{Machida2007,Machida2010,RauKee2014,Kavai2021} 

Atomic-resolution STEM confirms an abrupt, crystallographically coherent $\EIO/\TTO$ junction (Fig.~\ref{fig:fig1}b,c), preserving the pyrochlore framework across the interface and placing the two kagome-derived subsystems in atomic registry.\cite{Kareev2025} 
Since both $\TTO$ and $\DTO$ are insulating, the conducting channels in $\EIO/\TTO$ and $\EIO/\DTO$ remain associated with $\EIO$ (insets of Fig.~\ref{fig:fig1}d-f), whose low-temperature resistance shows the plateau-like behavior previously linked to the Fermi-arc transport.\cite{Wan2011,Resta2018,Armitage2018,Liu2021,Wu2025} We therefore retain the same itinerant Fermi-arc constituent while changing the local-moment environment presented to it.

The distinct interfacial coupling between Weyl Fermi arcs and $\TTO$ versus $\DTO$ is already visible at 20 mK and 18~T. While bare $\EIO$ shows only a weak intrinsic angular magnetoresistance (Fig.~\ref{fig:fig1}d), its coupling to either classical $\DTO$ or quantum $\TTO$ strongly reshapes the transport response (Fig.~\ref{fig:fig1}e,f). The earlier $\EIO/\DTO$ work established the classical limit, in which  predominantly longitudinal, dipolar spin-ice correlations are imprinted into the angular transport of $\EIO$.\cite{Wu2025,Zhang2023,Xing2025,Kareev2025} The $\EIO/\TTO$ interface does not simply reproduce that response. This high-field contrast motivates testing whether the same exchange-coupled architecture can encode quantum-spin-ice correlations, including Tb's transverse, multipolar degrees of freedom, into Weyl Fermi-arc transport.\cite{Onoda2011,Ruminy2016,Gingras2014,Rau2019}

\subsection*{Transducing the quantum-spin-ice correlations into Weyl Fermi-arc transport}
\begin{figure}[p]
\centering
\includegraphics[width=\textwidth]{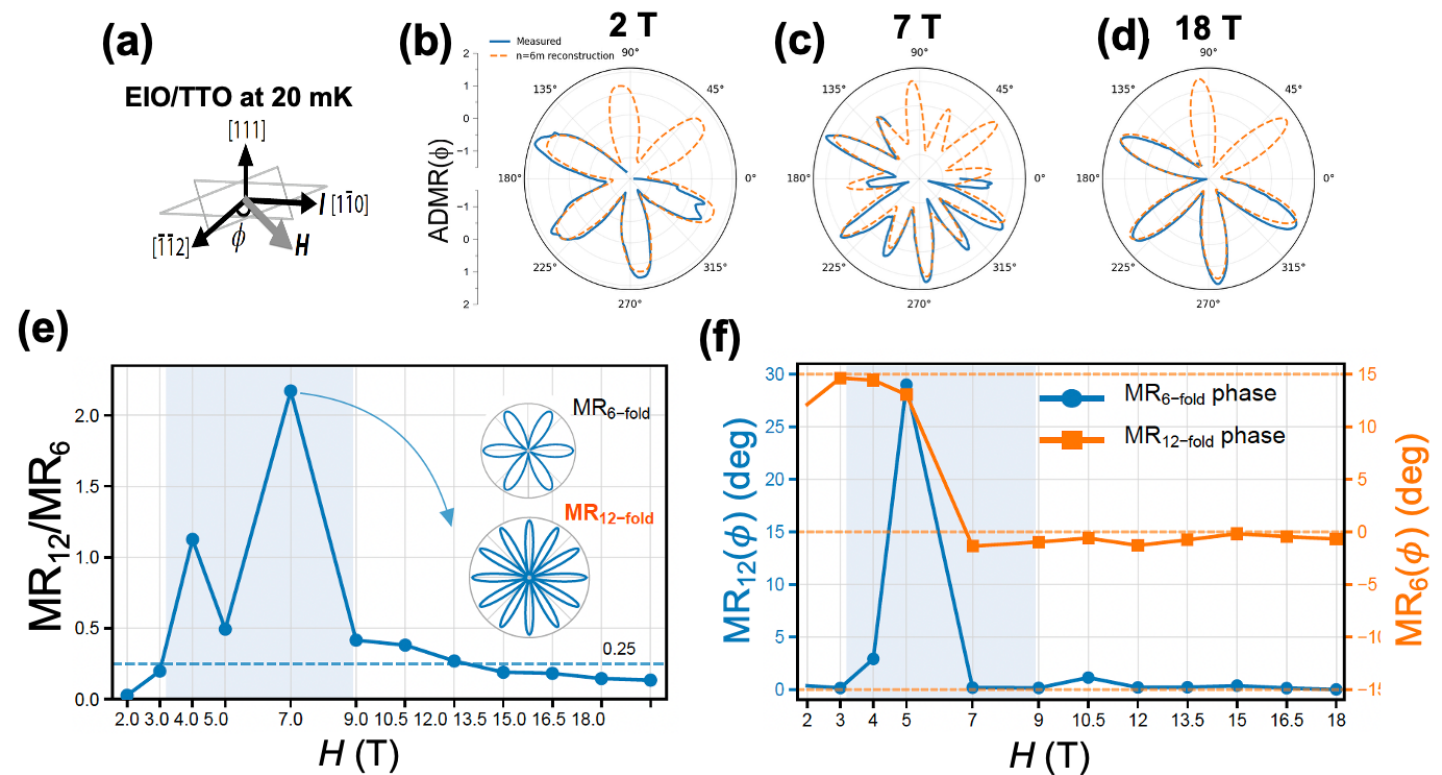}
\caption{\textbf{Transducing the quantum-spin-ice correlations into Weyl Fermi-arc transport.} \textbf{a}, In-plane field-rotation geometry for $\mathrm{Eu_2Ir_2O_7/Tb_2Ti_2O_7}$ at 20~mK. \textbf{b--d}, Representative angular magnetoresistance at 2, 7 and 18~T. Blue curves are measured data and orange dashed curves are harmonic reconstructions of the measured sector, illustrating the evolution from sixfold to twelvefold-rich and back to sixfold anisotropy. $\textrm{ADMR}(\phi)=[\rho_{xx}(\phi,H)-\rho_{xx}(0^{\circ},H)]/\rho_{xx}(0^{\circ},H)$, with $H$ rotating within the (111) plane. \textbf{e}, Relative harmonic weight $\mathrm{MR}_{12}/\mathrm{MR}_6$ versus field. The shaded interval highlights the intermediate-field reconstruction; the dashed line at 0.25 marks the minimal two-harmonic scale at which additional angular extrema become allowed (derivation and robustness tests in the Supplementary Information). Insets illustrate representative six- and twelvefold components. \textbf{f}, Field evolution of the orientations (phases) of the sixth- and twelfth-harmonic components. The same intermediate-field window exhibits a pronounced phase reorganization, showing that the reconstruction involves both harmonic weight and orientation. Phase-resolvability criteria and the full field series are given in the Supplementary Information.
}
\label{fig:fig2}
\end{figure}

To continuously tune the transverse, multipolar degrees of freedom, which are perpendicular to the local [111] orientation, we rotate the magnetic field within the (111) plane and alter the field magnitude while measuring the electronic transport response of $\EIO/\TTO$ (Fig.~\ref{fig:fig2}a). At 20~mK, the low-field response is dominated by six broad lobes with approximately $60^\circ$ repetition (Fig.~\ref{fig:fig2}b). On increasing the magnetic field strength, additional extrema appear between the original lobes, and the angular response reorganizes into an approximately $30^\circ$-periodic pattern near 7~T (Fig.~\ref{fig:fig2}c). At high field, the response returns to a predominantly sixfold morphology (Fig.~\ref{fig:fig2}d). The complete field evolution and its reproducibility across devices are shown in the Supplementary Information. The resulting sixfold$\rightarrow$twelvefold$\rightarrow$sixfold sequence is therefore a reentrant reconstruction of the electronic anisotropy, rather than a monotonic sharpening or polarization of a single low-field pattern.

The same reconstruction is evident in both the magnitude and the orientation of the symmetry-resolved response. The relative weight $\mathrm{MR}_{12}/\mathrm{MR}_6$ rises sharply from the low-field sixfold regime, becomes largest in the intermediate-field window, and falls again as sixfold dominance is recovered (Fig.~\ref{fig:fig2}e). The dashed reference marks the minimal two-harmonic scale at which an added twelfth harmonic can generate extra angular extrema; the derivation and robustness tests are given in the Supplementary Information. Independently, the phases of the sixth- and twelfth-harmonic components reorganize over the same field interval (Fig.~\ref{fig:fig2}f). Because harmonic weight and phase encode complementary parts of the complex angular response, their coincident evolution shows that the intermediate-field regime is a genuine reorganization of electronic symmetry and orientation.

\subsection*{Quantum and classical spin ice produce distinct symmetry-transduction trajectories}
\begin{figure}[p]
\centering
\includegraphics[width=\textwidth]{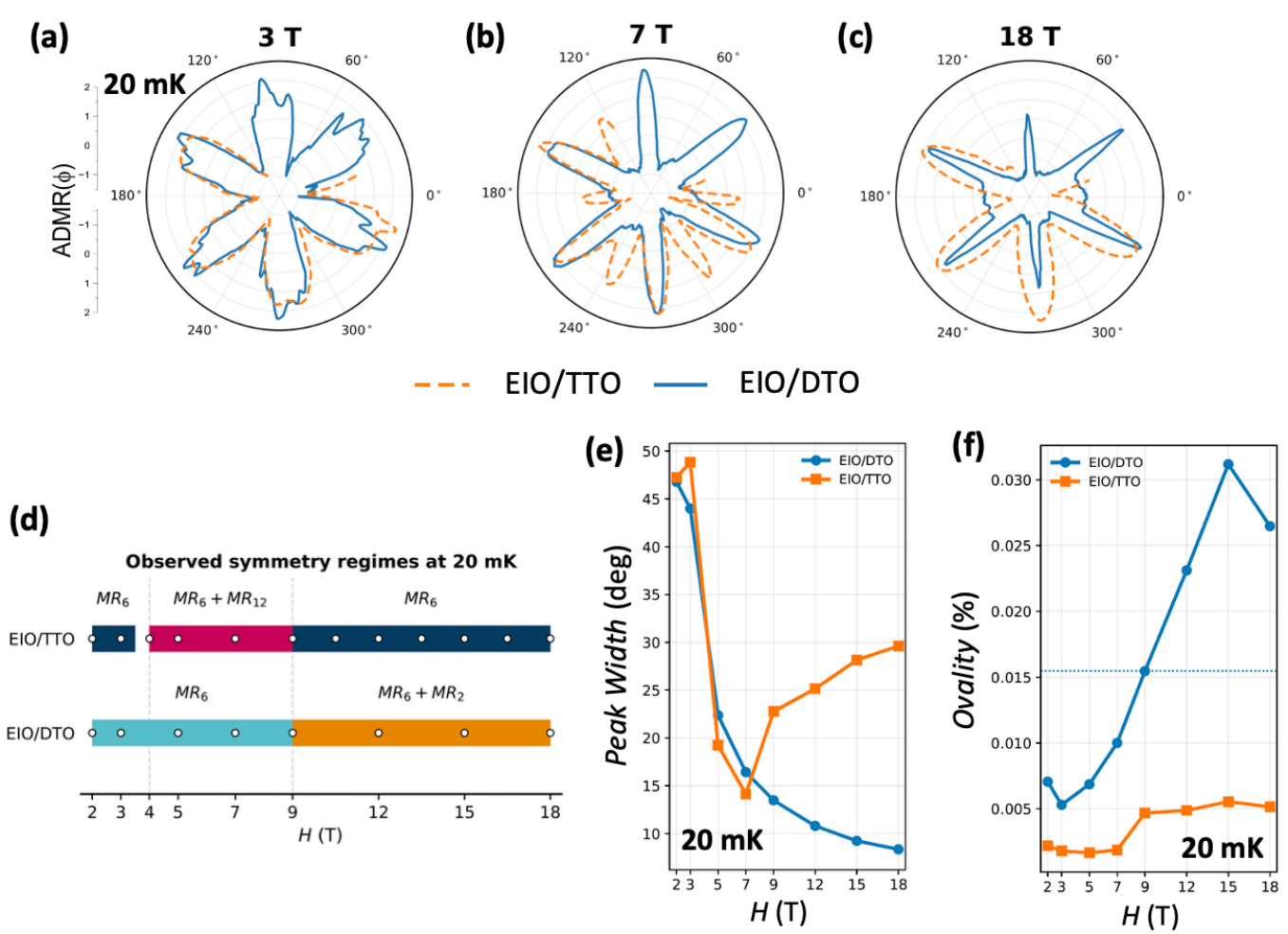}
\caption{\textbf{Quantum and classical spin ice produce distinct symmetry-transduction trajectories.} \textbf{a--c}, Angular magnetoresistance of $\mathrm{Eu_2Ir_2O_7/Tb_2Ti_2O_7}$ (orange dashed) and $\mathrm{Eu_2Ir_2O_7/Dy_2Ti_2O_7}$ (blue) at 20~mK and 3, 7 and 18~T. Only the $\mathrm{Tb_2Ti_2O_7}$-coupled interface develops the intermediate-field twelvefold reconstruction, whereas the high-field $\mathrm{Dy_2Ti_2O_7}$-coupled response sharpens and develops a pronounced twofold baseline distortion, consistent with the previously established classical-interface evolution.\cite{Wu2025,Zhang2023,Xing2025} \textbf{d}, Observed transport-symmetry regimes. Open circles mark measured fields; the narrow low-to-intermediate-field crossover in $\mathrm{Tb_2Ti_2O_7}$ is left unassigned. The ribbons summarize the dominant trajectories $\mathrm{MR}_6\rightarrow\mathrm{MR}_6+\mathrm{MR}_{12}\rightarrow\mathrm{MR}_6$ for $\mathrm{Eu_2Ir_2O_7/Tb_2Ti_2O_7}$ and $\mathrm{MR}_6\rightarrow\mathrm{MR}_6+\mathrm{MR}_2$ for $\mathrm{Eu_2Ir_2O_7/Dy_2Ti_2O_7}$ and do not denote thermodynamic phase boundaries. \textbf{e}, Mean local peak width at matched fields. The $\mathrm{Dy_2Ti_2O_7}$-coupled peaks narrow strongly with increasing field, whereas the $\mathrm{Tb_2Ti_2O_7}$-coupled response retains broad high-field lobes. \textbf{f}, Absolute polar-baseline ovality $e=(a-b)/(a+b)$, expressed as a percentage. The horizontal dotted line marks the $\mathrm{Dy_2Ti_2O_7}$ 9-T benchmark associated with the resolved high-field twofold contribution in the classical interface; the $\mathrm{Tb_2Ti_2O_7}$-coupled response remains below this scale throughout the measured range. The methods to acquire the mean peak width (\textbf{e}) and ovality (\textbf{f}) are given in the Supplementary Information.}
\label{fig:fig3}
\end{figure}

The origin of this reconstruction is sharpened by the classical-spin-ice control. Figures~\ref{fig:fig3}a--c compare $\mathrm{Eu_2Ir_2O_7/Tb_2Ti_2O_7}$ with $\mathrm{Eu_2Ir_2O_7/Dy_2Ti_2O_7}$ at matched fields while retaining the same $\mathrm{Eu_2Ir_2O_7}$ electronic constituent. $\mathrm{Dy_2Ti_2O_7}$ is the canonical predominantly classical dipolar spin ice, whereas $\mathrm{Tb_2Ti_2O_7}$ carries the additional non-Kramers transverse degrees of freedom discussed above.\cite{denHertog2000,Bramwell2001,Castelnovo2008,Gardner2010,Gingras2014,Rau2015,Onoda2011,Rau2019} At low field both interfaces are predominantly sixfold. Near 7~T, only $\mathrm{Eu_2Ir_2O_7/Tb_2Ti_2O_7}$ develops the additional extrema associated with the twelvefold response. At 18~T the two interfaces again differ qualitatively: the $\mathrm{Tb_2Ti_2O_7}$-coupled interface recovers broad sixfold lobes, whereas the $\mathrm{Dy_2Ti_2O_7}$-coupled interface evolves toward narrow sixfold peaks together with the twofold baseline distortion established previously for the classical interface.\cite{Wu2025,Zhang2023,Xing2025} The observed symmetry regimes therefore follow distinct trajectories (Fig.~\ref{fig:fig3}d):
\begin{equation*}
\begin{aligned}
\mathrm{Eu_2Ir_2O_7/Tb_2Ti_2O_7}:&\quad \mathrm{MR}_6\;\longrightarrow\;\mathrm{MR}_6+\mathrm{MR}_{12}\;\longrightarrow\;\mathrm{MR}_6,\\
\mathrm{Eu_2Ir_2O_7/Dy_2Ti_2O_7}:&\quad \mathrm{MR}_6\;\longrightarrow\;\mathrm{MR}_6+\mathrm{MR}_2.
\end{aligned}
\end{equation*}

Two independent geometric measures reinforce this quantum--classical separation. The characteristic peak widths begin on a comparable scale but diverge strongly with field: the $\mathrm{Dy_2Ti_2O_7}$-coupled features progressively sharpen, whereas the recovered $\mathrm{Tb_2Ti_2O_7}$-coupled sixfold lobes remain broad (Fig.~\ref{fig:fig3}e). The baseline ovality shows the complementary trend (Fig.~\ref{fig:fig3}f). The $\mathrm{Dy_2Ti_2O_7}$-coupled response develops a pronounced twofold deformation above the field scale identified in the earlier classical-interface study, while the $\mathrm{Tb_2Ti_2O_7}$-coupled response remains weakly oval over the same range.\cite{Wu2025,Zhang2023,Xing2025} Thus, the distinction between the two interfaces extends beyond the appearance of twelvefold extrema at one field to their entire field-evolution trajectories.

\subsection*{Discussion}
The $\EIO/\TTO$ and $\EIO/\DTO$ interfaces differ in the magnetic degrees of freedom being transduced, rather than in the mere presence of interfacial magnetic scattering. $\EIO/\DTO$ established that the field evolution of predominantly classical, longitudinal dipolar spin-ice correlations can be encoded into the anisotropic transport of the $\EIO$ electronic channel.\cite{Wu2025,Zhang2023,Xing2025} In $\TTO$, the non-Kramers doublet introduces transverse multipolar degrees of freedom in addition to the longitudinal Ising component.\cite{Onoda2011,Ruminy2016,Rau2019,Takatsu2016,Gritsenko2020,Kadowaki2022} The appearance of a twelvefold response only in $\EIO/\TTO$ therefore indicates that the electronic channel is sensitive to information associated with this enlarged quantum manifold.

A minimal two-sector framework couples the $\EIO$ electrons to longitudinal dipolar correlations and to transverse multipolar degrees of freedom. Their interfacial scattering profiles can have different angular forms and different field-dependent weights. We therefore write phenomenologically
\begin{equation}
\delta\rho_{\rm int}(\phi,H,T)=w_LF_L(\phi)+w_TF_T(\phi)+\lambda F_L(\phi)F_T(\phi)+\cdots,
\end{equation}
where $F_L$ and $F_T$ represent longitudinal and transverse correlation channels, respectively. 

The relevant field scale is also consistent with a reorganization of the low-energy Tb manifold. The first excited Tb crystal-field doublet lies approximately 1.5--1.6~meV above the ground doublet.\cite{Ruminy2016} For an illustrative Tb moment of $5.1\mu_B$ and the largest projection of a local $\langle111\rangle$ axis into the (111) field plane, the single-moment Zeeman scale is $E_Z^{\max}\simeq0.278B$~meV, reaching 1.5--1.6~meV near 5.4--5.7~T and spanning roughly 0.8--2.8~meV between 3 and 10~T.\cite{Sazonov2010,Sazonov2013} These energy scales, together with the known field sensitivity of the spin, magnetoelastic, and multipolar sectors,\cite{Rule2006,Cao2008,Ruff2010,Fennell2014,Ruminy2019,Turrini2021,Taniguchi2013,Takatsu2016,Roll2024} place the twelvefold transport reconstruction in the energy window where Tb correlations change substantially.

The atomically engineered frustrated-quantum-magnet/Weyl-semimetal heterostructures form a designer frustrated Kondo lattice platform in which the local moments and itinerant Weyl fermions are spatially separated yet exchange-coupled across an atomically coherent interface. This differs from conventional bulk Kondo lattices, where the local moments and conduction electrons are chemically intertwined;\cite{Machida2007,Machida2010,RauKee2014,Kavai2021} here, the magnetic manifold can be changed from the classical spin ice $\DTO$ to the quantum spin ice candidate $\TTO$ while retaining the same Weyl Fermi-arc transducer. Thus, the resulting difference in electronic anisotropy, especially the emergence of the twelvefold response at the quantum-spin-ice interface, illustrates how synthetic Kondo lattice interfaces can transduce not only classical correlations but also the enlarged quantum local-moment manifolds into Weyl Fermi-arc transport. 
\clearpage
\nolinenumbers
\begin{singlespace}\renewcommand*{\bibfont}{\small}
\printbibliography[title={References}]
\end{singlespace}
\clearpage
\clearpage
\clearpage
\clearpage

\linenumbers

\section*{Acknowledgements}
We acknowledge valuable discussions with J. Chakhalian, J. Pixley, Y. Chang, G. Malave, Y. Tang, A. Bodnar,  and M. J. Gingras. We also acknowledge the technical support from R. Nowell.

\section*{Funding}
T.-C.W., M.T., C.Z., and M.K. acknowledge the support by the U.S. Department of Energy, Office of Science, Office of Basic Energy Sciences under Award No. DE-SC0022160.  
This research is funded in part by a QuantEmX grant from ICAM and the Gordon and Betty Moore Foundation through Grant GBMF9616 to T.-C.W.
A portion of this work was performed at the National High Magnetic Field Laboratory, which is supported by National Science Foundation Cooperative Agreement No. DMR-1644779 and DMR-2128556 and the State of Florida. 
Q.Z. and L.G. acknowledge the support of the National Natural Science Foundation of China (52250402, 52025025).

\section*{Competing interests}
The authors declare no competing interests.

\end{document}